\documentclass[conference]{IEEEtran}
\IEEEoverridecommandlockouts

\usepackage[scr=rsfs]{mathalpha}
\usepackage{graphicx, amsfonts,amsmath , boldline, epstopdf,amssymb }
\usepackage{tabu, multirow,multicol ,booktabs ,setspace ,algcompatible ,mathrsfs,dsfont, array, boldline, makecell, booktabs }
\usepackage{soul}
\usepackage{amssymb}
\usepackage[linesnumbered,lined,boxed,commentsnumbered,ruled,longend]{algorithm2e}
\usepackage{adjustbox, stackengine, color, colortbl, epsfig,cite, cases, enumitem, mathtools, tikz, verbatim}
\usepackage{xcolor}
\definecolor{cpurple}{RGB}{112, 48, 160}
\definecolor{torange}{RGB}{190, 81, 8}
\definecolor{darkgreen}{RGB}{56,87,35} % Dark green color
\definecolor{lgreen}{RGB}{146,208,80} % Dark green color
\definecolor{wblue}{RGB}{0,112,192} % Dark green color
\definecolor{pred}{RGB}{192,0,0} % Dark green color

\usepackage{graphicx}
\usepackage{amsmath}

\newcolumntype{?}{!{\vrule width 1.2pt}} % for thicker the vertical line
\usepackage[caption = false]{subfig}
\usepackage[utf8]{inputenc}
\makeatletter

\begin{document}

\title{\vspace{-.1cm} Real-time Optimization for Wind-to-H$_2$ Driven Critical Infrastructures Based on Active Constraints Identification and Integer Variables Prediction \\
\thanks{This work is supported by U.S. National Science Foundation under Award CBET\#2124849.}
}

\author{\IEEEauthorblockN{ Mostafa Goodarzi}
\IEEEauthorblockA{\textit{Electrical and Computer Engineering} \\
\textit{University of Central Florida}\\
Orlando, USA \\
mostafa.goodarzi@ucf.edu}
\and
\IEEEauthorblockN{ Qifeng Li}
\IEEEauthorblockA{\textit{Electrical and Computer Engineering} \\
\textit{University of Central Florida}\\
Orlando, USA \\
Qifeng.Li@ucf.edu}}

\maketitle

\begin{abstract}
This paper proposes a concept of wind-to-hydrogen-driven critical infrastructure (W2H-CI) as an engineering solution for decarbonizing the power generation sector where it utilizes wind power to produce hydrogen through electrolysis and combines it with the carbon captured from fossil fuel power plants. First, a convex mathematical model of W2H-CI is developed. Then, an optimization model for optimal operation of W2H-CI, which is a large-scale mixed-integer convex program (MICP), is proposed. Moreover, we propose to solve this problem in real-time in order to hedge against the uncertainty of wind power. For this purpose, a novel solution method based on active constraints identification and integer variable prediction is introduced. This method can solve MICP problems very fast since it uses historical optimization data to predict the values of binary variables and a limited number of constraints which most likely contain all active constraints. We validate the effectiveness of the proposed fast solution method using two W2H-CI case studies.

%investigates the real-time optimal operation of the wind-to-hydrogen-driven critical infrastructure (W2H-CI) which is a new engineering solution for reducing carbon emissions from the power sector since it utilizes wind to produce hydrogen through electrolysis and combines it with the carbon captured from fossil fuel power plants. This paper first develops a convex mathematical model of W2H-CI, then proposes an optimization model for its real-time decision-making which is a mixed-integer convex program (MICP). Since our goal is to solve this problem in real-time, a novel solution method based on active constraints identification and integer variable prediction is introduced. This method can solve MICP problems very fast since it uses historical optimization data to predict the values of binary variables and a limited number of constraints which most likely contain all active constraints. We validate the effectiveness of the proposed fast solution method using two W2H-CI case studies.
\end{abstract}


\allowdisplaybreaks
\section{Introduction}   \label{sec: Intro}

Renewable energy sources, such as wind power, are widely recognized as key solutions for reducing carbon emissions and combating climate change. However, integrating wind power into electrical grids poses significant challenges, including hosting capacity limitations and grid stability issues like voltage fluctuations and flickers \cite{samet2020deep}. Wind-to-hydrogen (W2H) technology addresses these challenges by converting wind energy into hydrogen through water electrolysis, enabling efficient utilization of excess wind power while supporting grid stability. Recent studies have further enhanced the potential of W2H by integrating it with other systems to improve energy efficiency and carbon reduction. For example, W2H has been combined with carbon capture systems (CCS) to mitigate emissions \cite{samani2023optimal}, as well as with multi-energy networks, including electricity, thermal, and gas systems \cite{zhang2020environment}. Other studies explore its integration with components such as electric boilers, micro gas turbines, fuel cell (FC) charging stations \cite{li2021optimal}, and hydrogen refueling stations \cite{wu2020cooperative}, demonstrating the effectiveness of W2H in advancing sustainable energy systems.

This paper explores the integration of W2H technology with CCS into critical infrastructures (CIs), including water, power, hydrogen, and transportation, to achieve significant carbon emission reductions. Due to the substantial water demands of electrolysis, the water network is recognized as a crucial CI, an aspect often overlooked in previous W2H studies. Since water electrolysis connects to water distribution systems (WDS) and power networks at the distribution level, this study focuses on the distribution level. Additionally, the produced hydrogen can power FC vehicles (FCVs), incorporating transportation into the integrated system. This interconnected framework, known as W2H-driven critical infrastructures (W2H-CI), provides a comprehensive solution for reducing carbon emissions and strengthening the resilience of essential systems.

Some existing studies on W2H applications overlook wind uncertainty \cite{sun2021integration}, instead relying on day-ahead wind speed predictions, which may differ from actual data \cite{khezri2021demand}. Given the intermittent nature of wind, addressing uncertainty is critical for W2H applications. Although several studies have addressed wind energy uncertainty using probabilistic \cite{li2021optimal}, stochastic \cite{wu2020cooperative}, and robust optimization \cite{gu2019power} methods, these optimization models under uncertainty have significant limitations. For example, for large engineering systems, their deterministic approximations are often extremely large in scale and computationally challenging to solve. This paper proposes to solve the optimization of W2H-CI in real-time which does not rely on long-term (longer than 5 minutes) wind forecasts to hedge against uncertainty.

\textcolor{black}{We develop a convex mathematical model to formulate the W2H-CI problem as a real-time decision-making process,  which is a large-scale mixed-integer convex program (MICP). To solve such a computationally challenging optimization problem in real-time, we introduce a fast method based on the active constraints identification (ACI) and integer variable prediction (IVP) \cite{bertsimas2022online} that leverages supervised machine learning (SML) models. Based on historical data, the data-driven ACI-IVP method maps input parameters into the optimal values of binary variables and a set of active constraints. The key novelty lies in using ACI-IVP to surrogate our MICP problem with a small-scale continuous convex optimization problem that mature solvers, such as Mosek can rapidly solve. Our contributions include: (1) Integrating SML models with convex optimization for W2H-CI, (2) enabling real-time decision-making using ACI-IVP for intermittent wind energy, and (3) providing a scalable framework for practical CIs.}

The rest of this paper is organized as follows. Section \ref{sec: W2HCI} introduces the W2H-CIs and their modeling for remote areas with high penetration of wind energy. In Section \ref{sec: OpMeth}, we describe the methodology for real-time operation. Section \ref{sec: SolMeth} introduces the ACI-IVP method for solving the associated optimization problem. Section \ref{sec: CaseStudy} includes the numerical results and analysis. Finally, Section \ref{sec: Conclusions} presents the conclusions and suggests directions for future research.

\section{Wind-to-$H_{2}$-driven Critical Infrastructures}  \label{sec: W2HCI}
\begin{figure}[t]
% \vspace{-.5cm}
  \centering
{\includegraphics[width=.45\textwidth]{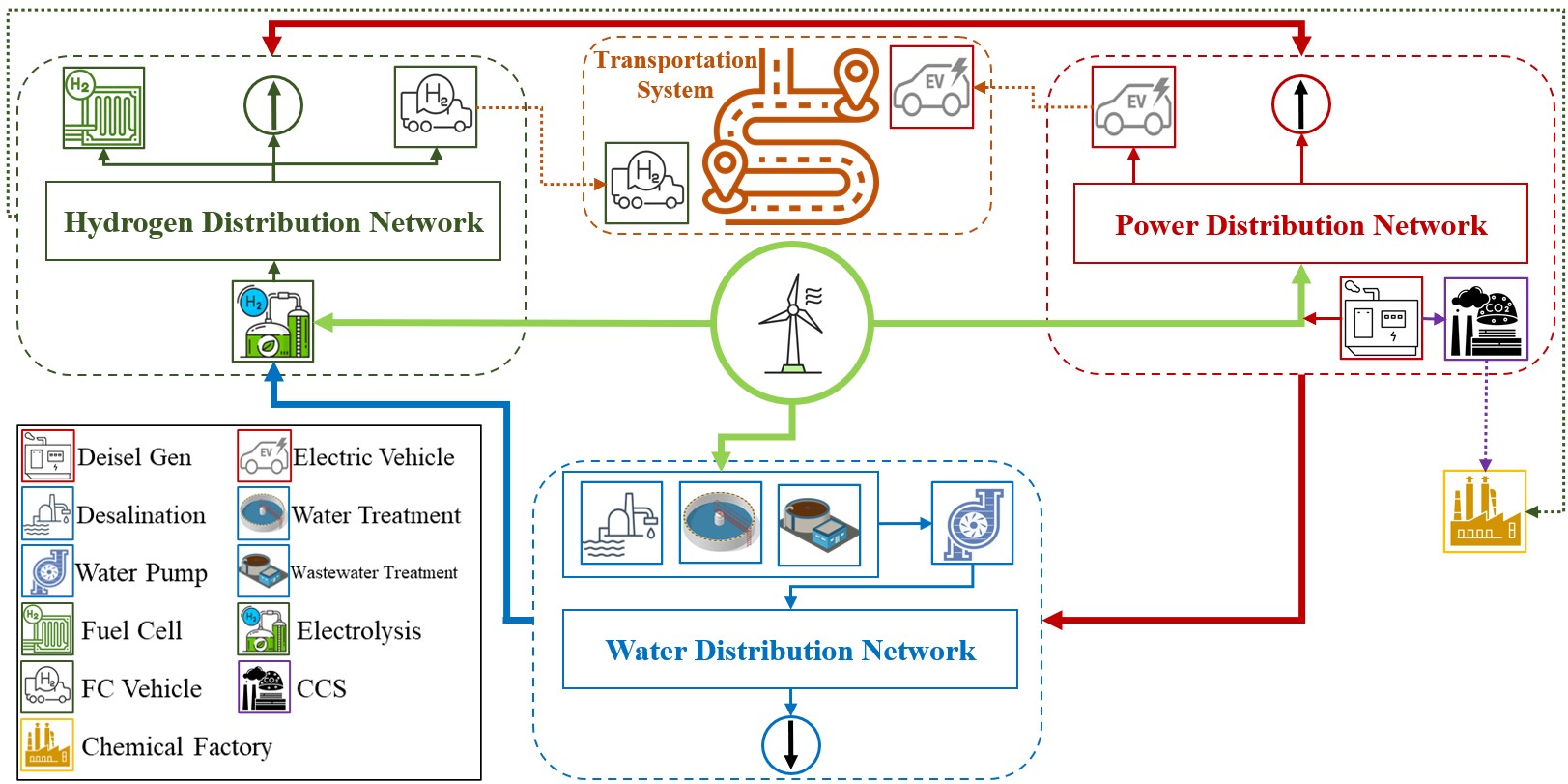}}
 \centering
           \vspace{-.3cm}
  \caption{\footnotesize The structure of W2H-CIs:  \textcolor{lgreen}{\textbf{$\rightarrow$} wind energy}, \textcolor{pred}{\textbf{$\rightarrow$} power}, \textcolor{wblue}{\textbf{$\rightarrow$} water}, \textcolor{darkgreen}{\textbf{$\rightarrow$} hydrogen}, \textcolor{cpurple}{\textbf{$\rightarrow$} carbon emission}, \textcolor{torange}{\textbf{$--$} transportation}.}
          \vspace{-.5cm}
  \label{Fig: W2H-CI}
\end{figure}
%%%%%%%%%%%%%%%%%%%%%%%%%%%%%%%%%%%

This section explains the different components and the mathematical model of a W2H-CI. Fig. \ref{Fig: W2H-CI} shows a typical W2H-CI that includes a power section (red), a water section (blue), a hydrogen section (green), and a transportation section (orange). This W2H-CI is particularly well-suited for remote coastal cities and small islands, where independent local power systems and abundant wind energy are available. A wind farm provides the energy needs of all CIs in a specific remote coastal city or small island, where the part of wind power that is within the hosting capacity will be directly fed into the power distribution system. The rest portion, i.e., the volatile part, of wind power will be used for desalination and electrolyzing water. The power system has a diesel generator equipped with a CCS and a FC providing backup during periods of limited wind energy availability. The WDS is connected to the power system for its energy and to the hydrogen system to supply water for electrolysis. Hydrogen generated by electrolysis is used in various applications, including methanation to reuse captured carbon, converting unpredictable wind energy into controllable energy through FC units, and satisfying the hydrogen network's demand like FCVs in the transportation system. In the following subsections, we explain the convex mathematical models of different components of the W2H-CI. The transportation system is considered in the power and hydrogen sections. For further details on the convex-hull formulation and its accuracy, please refer to \cite{li2018micro,goodarzi2023hybrid}.

\subsection{Power Section} \label{sec:PFPower} 
Power flow in the power distribution network (PDN) can be modeled using several formulations. The Distflow model involves bus variables and branch variables and can be used to model both active and reactive power flow in the PDN. The convex model of the power section is expressed by \cite{li2017convex}:
\begin{subequations} \label{eq_PDNmodel}
	\begin{align}
 &\resizebox{.88\hsize}{!}{$\sum\limits_{k}(p_{ki,t})+r_{ij}{\mathcal{I}}_{ij,t}-p_{ij,t}=p^\mathrm{dg}_{i,t}+p^\mathrm{sw}_{i,t}+p^\mathrm{hs}_{i,t}+p^\mathrm{ts}_{i,t}-p^\mathrm{l}_{i,t}$}, \label{eq_PDN1}\\
&\resizebox{.88\hsize}{!}{$\sum\limits_{k}(q_{ki,t})+x_{ij}{\mathcal{I}}_{ij,t}-q_{ij,t}=q^\mathrm{dg}_{i,t}+q^\mathrm{sw}_{i,t}+q^\mathrm{hs}_{i,t}+q^\mathrm{ts}_{i,t}-q^\mathrm{l}_{i,t}$}, \label{eq_PDN2}\\
&\mathcal{V}_{i,t}-\mathcal{V}_{j,t}=2(r_{ij}p_{ij,t}+x_{ij}q_{ij,t})-(r^\mathrm{2}_{ij}+x^\mathrm{2}_{ij}) {\mathcal{I}}_{ij,t}, \label{eq_PDN3}\\
&p_{ij,t}^2+q_{ij,t}^2 \leq \mathcal{V}_{i,t}{\mathcal{I}}_{ij,t},\label{eq_PDN4}\\
&\underline{\mathcal{V}}_{i} \overline{\mathcal{V}}_{i} {\mathcal{I}}_{ij,t} + {\overline{s}_{ij}}^2 \mathcal{V}_{i,t} \leq {\overline{s}_{ij}}^2 (\underline{\mathcal{V}}_{i} + \overline{\mathcal{V}}_{i})\label{eq_PDN4_2}\\
&p_{ij,t}^2+q_{ij,t}^2\leq {\overline{s}_{ij}}^2,\label{eq_PDN5}\\
&\resizebox{.88\hsize}{!}{$\underline{p}^\mathrm{dg}_{i},\underline{q}^\mathrm{dg}_{i},\underline{\mathcal{V}}_{i},\underline{\mathcal{I}}_{ij} \leq p^\mathrm{dg}_{i,t},q^\mathrm{dg}_{i,t},\mathcal{V}_{i,t},\mathcal{I}_{ij,t}\leq \overline{p}^\mathrm{dg}_{i},\overline{q}^\mathrm{dg}_{i},\overline{\mathcal{V}}_{i},\overline{\mathcal{I}}_{ij}$}, \label{eq_PDN6}\\
% &\underline{P}_{i},\underline{Q}_{i},\underline{\mathcal{V}}_{i},\underline{\mathcal{I}}_{ij} \leq P_{i,t},Q_{i,t},\mathcal{V}_{i,t},\mathcal{I}_{ij,t}\leq \overline{P}_{i},\overline{Q}_{i},\overline{\mathcal{V}}_{i},\overline{\mathcal{I}}_{ij}\label{eq_PDN6}\\
%%%%%%%%%%%%%%%%%%%%%%%%%%%%%%%%%%%% NET POWER
&p^\textrm{sw}_t = p^\textrm{wind}_t - (p^\textrm{we}_t + p^\textrm{wd}_t), \label{eq_netPower}\\
% %%%%%%%%%%%%%%%%%%%%%%%%%%%%%%%%%%%%%%%% C emitted
&c_t^\textrm{dg} = \xi^\textrm{dg}_\textrm{c}p^\textrm{dg}_t,\label{eq_CCS_1}\\
&c^\textrm{e}_t = c_t^\textrm{dg} -  c^{\chi}_t,\label{eq_CCS_2}
	\end{align}
\end{subequations}
where $k$, $i$, and $j$ denote bus numbers, and $t$ represents time. $\overline{s}$, $\mathcal{I}$, $p$ and $q$, and $r$ and $x$ represent the maximum apparent power, square current, active and reactive power flow, and resistance and reactance of the branches, respectively. $\mathcal{V}$, $p^\mathrm{dg}$, $q^\mathrm{dg}$, $p^\mathrm{sw}$, $q^\mathrm{sw}$, $p^\mathrm{hs}$, $q^\mathrm{hs}$, $p^\mathrm{ts}$, $q^\mathrm{ts}$, $p^\mathrm{l}$, and $q^\mathrm{l}$ are the square voltage, active and reactive power of diesel generation, surplus wind, hydrogen system, transportation system, and power load at each bus, respectively. The nodal balance of active and reactive power can be determined by (\ref{eq_PDN1}) and (\ref{eq_PDN2}), respectively. Constraints (\ref{eq_PDN3}) to (\ref{eq_PDN5}) are related to Ohm's law. The upper and lower bounds for variables are described by (\ref{eq_PDN6}).
$p^\mathrm{wind}$, $p^\mathrm{we}$, and $p^\mathrm{wd}$ in (\ref{eq_netPower}) are the total power of wind energy, water electrolysis, and water desalination. In cases where the diesel generator is used as a backup, carbon emissions are decreased with a CCS, as illustrated in (\ref{eq_CCS_1}) and (\ref{eq_CCS_2}). The carbon emissions from diesel generators are represented by $c_t^\textrm{dg}$, a portion of which is captured for reuse ($c^{\chi}$), while the remainder is emitted ($c^\textrm{e}$).

\subsection{Water Section} \label{sec:PFWater}
A convex-hull model for the water section of the W2H-CI, including water desalination, mass flow conservation law, pipe flow, water pumps, and water tank, is given \cite{goodarzi2022evaluate,li2018micro,goodarzi2023hybrid}: 
\begin{subequations} \label{eq_WDNmodel}
	\begin{align}
&\sum\limits_{m}f_{nm,t} = f^\mathrm{wd}_{n,t}-d_{n,t}+f^\mathrm{wt}_{n,t}, \label{eq_WDN1}\\
   &y_{n,t}-y_{m,t}+h_{nm} = \nonumber \\
&\begin{cases}
      \leq (2\sqrt{2}-2)r^\mathrm{w}_{q}\overline{f}_{q}f_{q,t}+(3-2\sqrt{2})r^\mathrm{w}_{q}\overline{f}_{q}^2, \\
      \geq (2\sqrt{2}-2)r^\mathrm{w}_{q}\underline{f}_{q}f_{q,t}-(3-2\sqrt{2})r^\mathrm{w}_{q}\underline{f}_{q}^2,  \\
      \geq 2r^\mathrm{w}_{q}\overline{f}_{q}f_{q,t}-r^\mathrm{w}_{q}\overline{f}_{q}^2,  \\
      \leq 2r^\mathrm{w}_{q}\underline{f}_{q}f_{q,t}+r^\mathrm{w}_{q}\underline{f}_{q}^2.
    \end{cases} \label{eq_WDN2}\\
&\resizebox{.88\hsize}{!}{$y_{n,t}-y_{m,t}+h_{nm} +y^{\mathrm{G}}_{q,t}-r^\mathrm{w}_{q}(f_{q,t})^2 \ge M(b^\mathrm{p}_{q,t}-1)$},\label{Convwatp1}\\
&\resizebox{.88\hsize}{!}{$y_{n,t}-y_{m,t}+h_{nm} +y^{\mathrm{G}}_{q,t}-r^\mathrm{w}_{q}\overline{f}_{q}f_{q,t} \le M(1-b^\mathrm{p}_{q,t})$},\label{Convwatp2}\\
 &0 \le f_{nm} \le b^\mathrm{p}_{p,t} \overline{f}_{p}.\label{Convwatp3}\\
&V_{n,t+1}^\mathrm{wt}=V_{n,t}^\mathrm{wt}+f_{n,t}^\mathrm{wt},  \label{eq_WDN3}\\
&\underline{f}^\mathrm{wd}_{n}, \underline{f}^\mathrm{wt}_{n},\underline{V}^\mathrm{wt}_{n} \leq f^\mathrm{wd}_{n,t}, f^\mathrm{wt}_{n,t},V^\mathrm{wt}_{n,t} \leq \overline{f}^\mathrm{wd}_{n}, \overline{f}^\mathrm{wt}_{n},\overline{V}^\mathrm{wt}_{n},\label{eq_WDN4}\\
% &\underline{y}_{n}, \underline{f}_{p} \leq y_{n,t}, f_{p,t},\leq \overline{y}_{n}, \overline{f}_{p},\label{eq_WDN5}\\
% %%%%%%%%%%%%%%%%%%%%%%%%%%%%%%%%%%%% P desalination and pump
&\resizebox{.88\hsize}{!}{$p^\mathrm{wd}_t = b^\mathrm{wd}_t e^\mathrm{wd}_\mu  f^\textrm{wd}_t, \,\, 0.25(\mu-1) \overline{f}^\textrm{wd} \leq f^\textrm{wd}_t \leq 0.25\mu \overline{f}^\textrm{wd}$}, \label{eq_WNP1}\\
&\eta p^\mathrm{p}_{i,t} \ge 2.725\times (a_1(f_{q,t})^2+a_0f_{q,t})\label{ConvPump_1}\\
&\eta p^\mathrm{p}_{i,t} \le 2.725\times (a_1\overline f_{q}+a_0) f_{q,t}.\label{ConvPump_2}
%%%%%%%%%%%%%%%%%%%%%%%%%%%%%%%%%%%%%%%%%%%%%%%%%%%%%%%%%%
\end{align}
\end{subequations}
\textcolor{black}{Where $n$, $m$, and $q$ denote the number of nodes and pipes, respectively. $f$, $d$, $y$, $h$, and $r^\mathrm{w}$ represent water flow, water demand, water head pressure, elevation, and the pipe's head loss coefficient, respectively. The variables $f^\mathrm{wd}$, $f^\mathrm{wt}$, $y^{\mathrm{G}}$, $V^\mathrm{wt}$, and $p^\mathrm{p}$ denote the water flow from desalination, the water flow from the tank, the head gain of the pump, the volume of the water tank, and the pump power, respectively. The binary variables $b^\mathrm{p}$ and $b^\mathrm{wd}$ represent the on/off status of the pump and water desalination. $\eta$ and $e$ are constant parameters. The equality of water injection and water output at each node is ensured by (\ref{eq_WDN1}). Constraint (\ref{eq_WDN2}) defines the convex-hull model for head loss in a regular pipe, while the convex model for a pipe with a pump is captured by (\ref{Convwatp1}) to (\ref{Convwatp3}). Each tank is modeled as a node using (\ref{eq_WDN3}), and the convex model of a desalination and a pump are described by (\ref{eq_WNP1}) and (\ref{ConvPump_2}).}

\subsection{Hydrogen Section}\label{sec:Hydrogen}
The following mathematical formulations describe the hydrogen section of the W2H-CI, which includes water electrolysis, an FC unit, and a hydrogen tank:
\begin{subequations} \label{eq_WHNmodel}
	\begin{align}
&h^\textrm{we}_t = b^\textrm{we}_t \xi^\textrm{we}_\textrm{p} p^\textrm{we}_t \label{eq_H2_1}\\
&d^\textrm{we}_t = b^\textrm{we}_t \xi^\textrm{we}_\textrm{w}  h^\textrm{we}_t,\label{eq_H2_2}\\
&p^\textrm{fc}_t =  b^\textrm{fc}_t \xi^\textrm{fc}_\textrm{h} h^\textrm{fc}_t,\label{eq_H2_3}\\
&(p^\textrm{we}_t - p^\textrm{fc}_t)^2 + (q^\textrm{hs}_t)^2 \leq (\overline{s}^\textrm{hs})^2,\label{eq_H2_4}\\
&V^\textrm{ht}_{t+1} = (1-\xi^\textrm{dsp})V^\textrm{ht}_{t} + (h^\textrm{we}_t - h^\textrm{fc}_t - h^\textrm{ts}_t),\label{eq_H2_5}\\
&b^\textrm{we}_t + b^\textrm{fc}_t \leq 1,\label{eq_H2_6}\\
&\underline{h}^\textrm{fc}, \underline{h}^\textrm{we}, \underline{p}^\textrm{we} \le h^\textrm{fc}_t, h^\textrm{we}_t, {p}^\textrm{we}_t \leq \overline{h}^\textrm{fc}, \overline{h}^\textrm{we}, \overline{p}^\textrm{we},\label{eq_H2_7}
\end{align}
\end{subequations}
where $h^\textrm{we}$, $h^\textrm{ts}$, and $h^\textrm{fc}$ denote the amount of hydrogen produced by electrolysis and utilized by the transportation system and FC system, respectively. $\xi^\textrm{we}_\textrm{p}$, $\xi^\textrm{we}_\textrm{w}$, $\xi^\textrm{fc}_\textrm{h}$, and $\xi^\textrm{dis}$ are energy for $H_2$ production factor, water for $H_2$ production factor, $H_2$ power generation factor, and hydrogen tank disappear factor. The power associated with electrolysis and the FC system is represented by $p^\textrm{we}$ and $p^\textrm{fc}$, while $b^\textrm{we}$ and $b^\textrm{fc}$ are binary variables related to the operation of the electrolysis and FC.

% %%%%%%%%%%%%%%%%%%%%%%%%%%%%%%%%%%%%%%% H System
% &h^\textrm{we}_t = b^\textrm{we}_t \xi^\textrm{we}_\textrm{p} p^\textrm{we}_t \label{eq_H2_1}\\
% &d^\textrm{we}_t = b^\textrm{we}_t \xi^\textrm{we}_\textrm{w}  h^\textrm{we}_t,\label{eq_H2_2}\\
% &p^\textrm{fc}_t =  b^\textrm{fc}_t \xi^\textrm{fc}_\textrm{h} h^\textrm{fc}_t,\label{eq_H2_3}\\
% &(p^\textrm{we}_t - p^\textrm{fc}_t)^2 + (q^\textrm{hs}_t)^2 \leq (\overline{s}^\textrm{hs})^2,\label{eq_H2_4}\\
% &V^\textrm{ht}_{t+1} = (1-\xi^\textrm{dsp})V^\textrm{ht}_{t} + (h^\textrm{we}_t - h^\textrm{fc}_t - h^\textrm{ts}_t - h^\textrm{zb}_t - h^\textrm{if}_t),\label{eq_H2_5}\\
% &b^\textrm{we}_t + b^\textrm{fc}_t \leq 1,\label{eq_H2_6}\\
% &\underline{h}^\textrm{fc}, \underline{h}^\textrm{we}, \underline{p}^\textrm{we} \le h^\textrm{fc}_t, h^\textrm{we}_t, {p}^\textrm{we}_t \leq \overline{h}^\textrm{fc}, \overline{h}^\textrm{we}, \overline{p}^\textrm{we},\label{eq_H2_7}\\
% %%%%%%%%%%%%%%%%%%%%%%%%%%%%%%%%%%%%%%%% C emitted
% &c_t^\textrm{dg} = \xi^\textrm{dg}_\textrm{c}p^\textrm{dg}_t,\label{eq_CCS_1}\\
% &c^\textrm{e}_t = c_t^\textrm{dg} -  c^{\chi}_t.\label{eq_CCS_2}
% 	\end{align}
% \end{subequations}

\section{Optimization Models for real-time operation}  \label{sec: OpMeth}
The proposed W2H-CIs rely on actual wind speed values to address uncertainties and ensure optimal real-time operation. The objective function for this optimization model aims to minimize carbon emissions and the cost associated with using diesel generators, represented by:
\begin{align} \label{eq_Objective}
\sum_{t=1}^{T} a_1 p^\mathrm{dg}_{t} + a_2 c_t^\textrm{e}, 
\end{align}
where $a_1$ to $a_2$ are penalty \cite{shao2019low} or cost parameters. In this paper, we consider that 5-minute-ahead wind speed predictions provide sufficient accuracy for real-time operations. The operational flexibility of devices such as FCs, desalination plants, and water pumps enables rapid changes in their status \cite{goodarzi2022evaluate}, simplifying their control. Similarly, PEM electrolysis, known for its fast response\cite{daud2017pem}, is also straightforward to manage.
Given the inclusion of water and hydrogen tanks capable of storing and utilizing resources throughout the day, a 24-hour operational horizon is proposed. To achieve effective real-time control, we recommend employing a rolling window approach, solving the optimization problem at 5-minute intervals over 24 hours. The comprehensive optimization model for W2H-CI is formulated as follows:
\begin{align} \label{eq: OpMo}
& \textbf{min} \,\,\,\,\ (\ref{eq_Objective}) \nonumber  \\
& s.t \,\,\,\,\,\,\,\ (\ref{eq_PDN1}) - (\ref{eq_CCS_2}) \nonumber  \\
& \,\,\,\,\,\,\,\,\,\,\,\,\,\,\,\, \mathcal{B} \in \{0,1\},
\end{align}
where $\mathcal{B}$ is the set of binary variables regarding pump, electrolysis, desalination, and FC.

\section{The ACI-IVP-Based Solution Method}  \label{sec: SolMeth}
% \section{Predictive Surrogate for Continuous Relaxation Method}
Computational efficiency is critical to achieving real-time solutions for the problem (\ref{eq: OpMo}). Despite convexifying the non-convex constraints, the presence of binary variables and the large-scale nature of the problem still impose significant computational challenges when using conventional optimization techniques. To overcome these challenges, we present the surrogate transformation method, a novel approach based on ACI-IVP. This method is specifically designed to address the computational demands of solving MICP problems \cite{bertsimas2023prescriptive}, enabling real-time optimal operation for W2H-CIs.

The ACI-IVP-based surrogate transformation method utilizes machine learning to simplify the optimization problem while preserving solution accuracy. First, it identifies the active constraints using SML techniques such as DT, KNN, NB, and SVM. Active constraints are defined as the set of constraints satisfied as equalities at the optimal solution. Non-active constraints, which do not influence the solution, are filtered out. This process significantly reduces the complexity of the problem by transforming the original problem, which contains $M$ constraints, into a simplified surrogate problem with only $n$ active constraints, where $n << M$. Next, the method predicts the optimal values of binary variables using SML techniques and substitutes these values into the original problem. This step eliminates all integer variables, transforming the problem into a purely continuous optimization problem.
Once the non-active constraints are filtered out and the binary variables are replaced with their predicted optimal values, the original problem is transformed into a simplified surrogate problem. This surrogate problem, free of binary variables and non-active constraints, becomes significantly easier and faster to solve using optimization solvers. 
% These solvers can compute the solution rapidly due to the substantially reduced complexity of the problem. 
% Since problem (\ref{eq: OpMo}) needs to be solved in real-time, achieving computational efficiency is essential for the solution method. Despite convexifying these non-convex constraints, the presence of binary variables and the large problem size continue to pose significant computational challenges when using conventional optimization techniques. This section presents the surrogate transformation method, an ACI-IVP-based approach specifically developed to quickly solve MICP problems for W2H-CI, making it well-suited for achieving real-time optimal solutions \cite{bertsimas2023prescriptive}. Fig. \ref{pic: CST} illustrates the surrogate transformation process to obtain the optimal solution. The surrogate transformation method filters the non-active constraints and predicts the optimal values of binary variables. It results in a simplified surrogate problem that contains only active constraints ($n << M$) and continuous variables. The reduced problem can be solved using standard solvers such as MOSEK or Gurobi.
\begin{algorithm}[t!]
\footnotesize
\SetAlgoLined
\KwIn{Real-time wind speed, power, water, and hydrogen demand}
\KwOut{Optimal operational of W2H-CI}

Find the optimal values for binary variables and identify active constraints using the ACI-IVP-based filtering method and equation (\ref{eq_map})\;
Update (\ref{Convwatp1})-(\ref{Convwatp3}), (\ref{eq_WNP1}), (\ref{eq_H2_1})-(\ref{eq_H2_3}) with optimal values of binary variables, and remove (\ref{eq_H2_6})\;
Identify active constraints and filter out redundant constraints\;
Develop a surrogate optimization problem\;
Solve the surrogate optimization problem using optimization solver\;

\caption{\footnotesize Solving the MICP of the W2H-CI in real-time}
\label{alg: CSTmeth}
\end{algorithm}
\begin{figure}[!t]  
    \vspace{-.35cm}
  \centering
\includegraphics[width=0.375\textwidth]{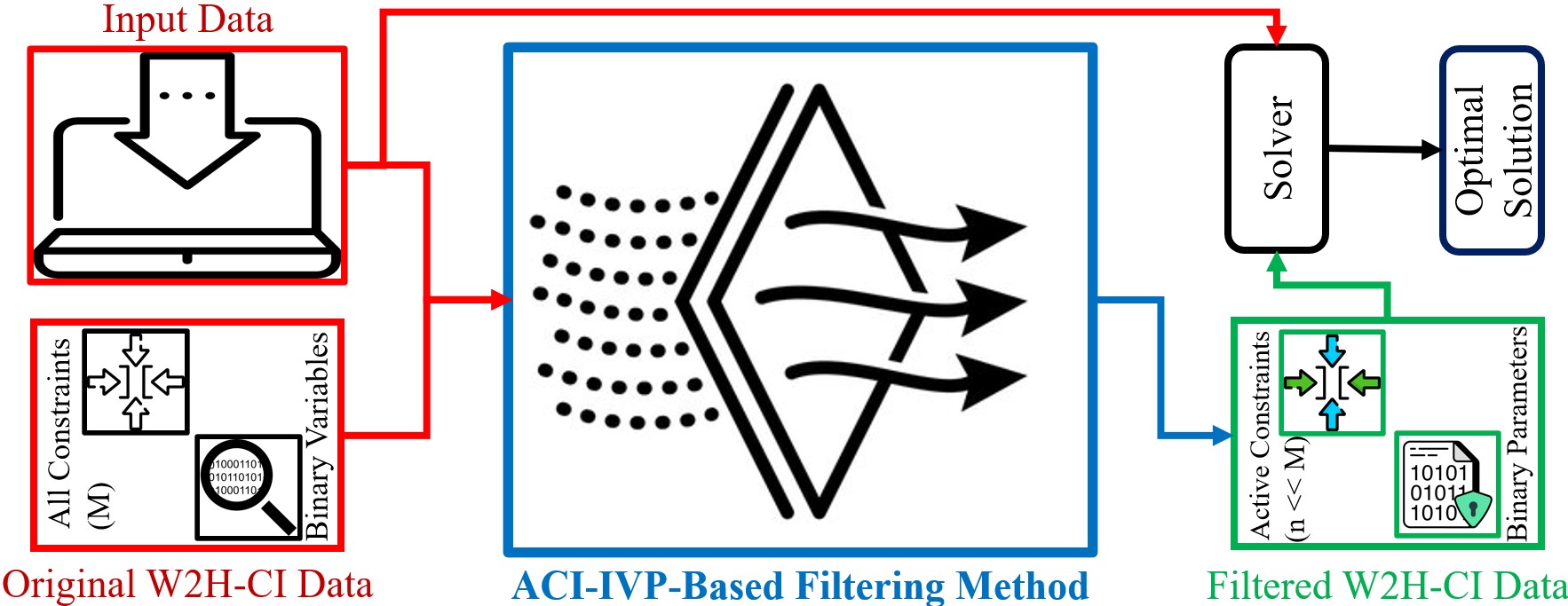}
 \centering
     \vspace{-.25cm}
    \caption{ACI-IVP-based surrogate transformation method}
  \label{pic: CST} 
  \vspace{-.55cm}
\end{figure} 
% \begin{figure}[!t]  
%     % \vspace{-.55cm}
%   \centering
% \includegraphics[width=0.3\textwidth]{Figs/CST 2.jpg}
%  \centering
%      \vspace{-.3cm}
%     \caption{-}
%   \label{pic: CST} 
%   \vspace{-.6cm}
% \end{figure} 

The data-driven filtering method within the ACI-IVP-based surrogate transformation process is a SML model that maps input data—including real-time values of wind speed, water, power, and hydrogen demand of CIs to outputs such as optimal binary variables and active constraints.
\begin{equation} \label{eq_map} 
\vartheta = \mathcal{H}(\varphi), \end{equation}
where $\varphi$ and $\vartheta$ represent the input and output of the prediction model, respectively. The model is trained using the dataset $\Gamma =\{(\varphi_1,\vartheta_1),(\varphi_2,\vartheta_2),...,(\varphi_{\mathcal{N}},\vartheta_{\mathcal{N}})\}$, which is generated through the solving of offline optimization problems. An accurate hypothesis function directly depends on the quality of the training dataset, so it is vital to build a reliable, robust dataset. Algorithm \ref{alg: CSTmeth} and Fig. \ref{pic: CST} illustrate the ACI-IVP-based surrogate transformation method for real-time optimal operation of the W2H-CIs.

\section{Case Studies}  \label{sec: CaseStudy}
\begin{figure}[!b]  
    \vspace{-.6cm}
  \centering
\includegraphics[width=0.35\textwidth]{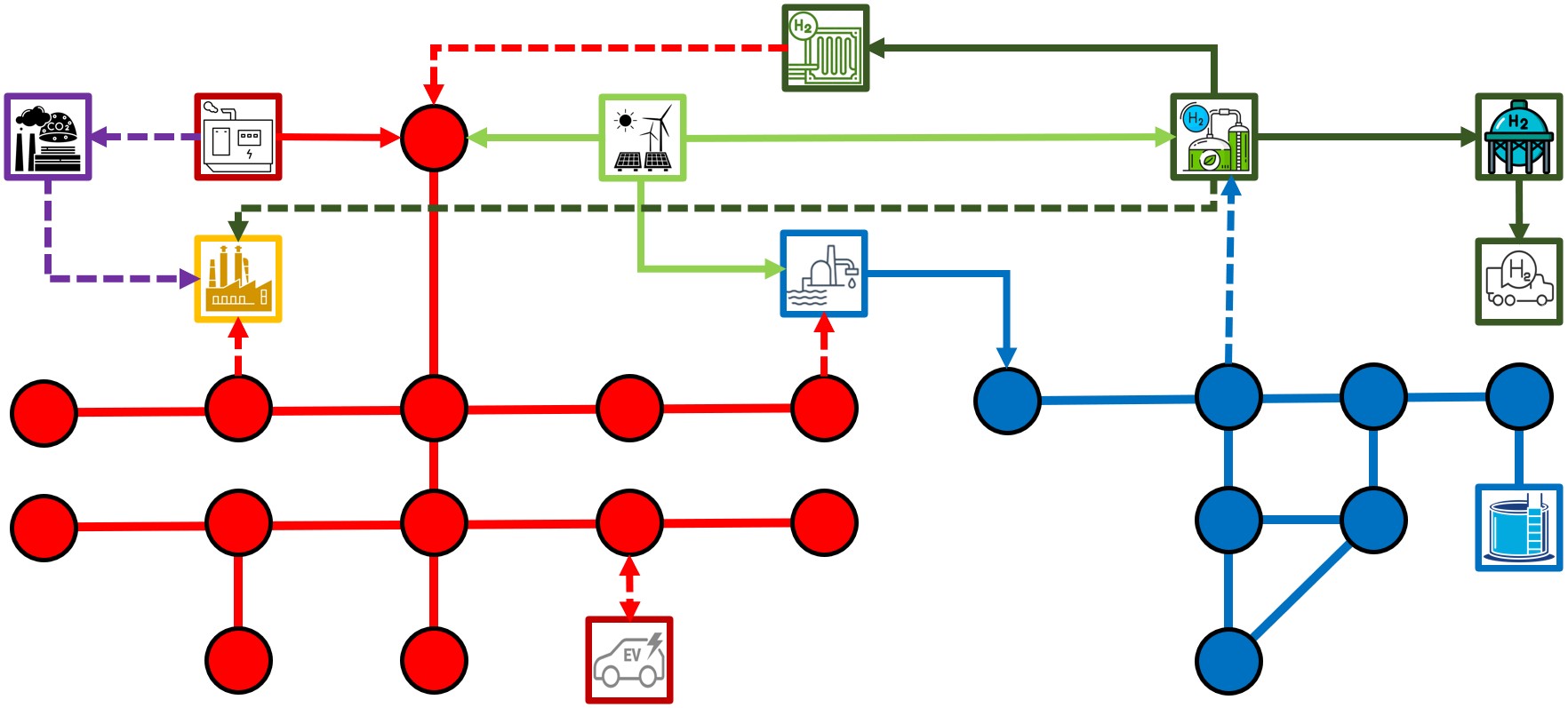}
 \centering
 \vspace{-.3cm}
    \caption{W2H-CIs for a small community such as industrial parks.}
  \label{pic: IEEE13}   
      % \vspace{-.2cm}
\end{figure}

\begin{figure}[!t]  
    % \vspace{-.7cm}
  \centering
\includegraphics[width=0.35\textwidth]{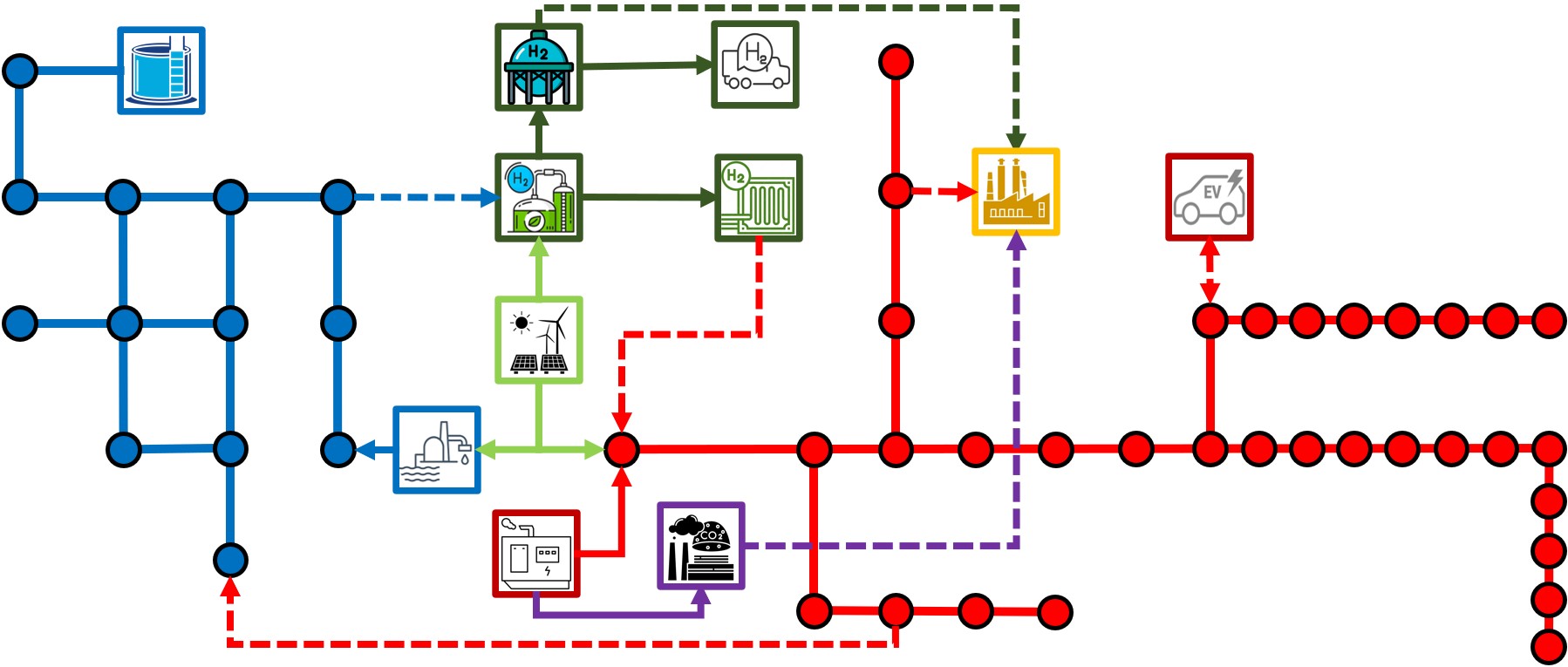}
 \centering
 \vspace{-.3cm}
    \caption{W2H-CIs for city-scale application.}
  \label{pic: IEEE33}   
      \vspace{-.2cm}
\end{figure}

\textcolor{black}{Based on the characterization of CIs at the distribution level \cite{gilvanejad2021introduction}, the robustness of the proposed method is demonstrated using two test systems shown in Fig. \ref{pic: IEEE13} and Fig. \ref{pic: IEEE33}.}
\textcolor{black}{In these figures, red, blue, light green, dark green, and purple, represent the power, water, wind energy, hydrogen, and carbon emission of W2H-CI, respectively. The transportation system is integrated with power and hydrogen sections to support FCVs and electric vehicles.
The first test case involves a modified IEEE 13-bus system integrated with an 8-node EPANET water system, representing small community W2H-CIs such as industrial parks. The second test case uses an IEEE 33-bus system paired with a 13-node Otsfeld water system \cite{goodarzi2022security}, modeling city-scale W2H-CIs. In larger-scale systems, CIs function independently, leading to conflicts between them. Because of these conflicts, the proposed method, though mathematically feasible, cannot currently be used.
The training datasets for this study are generated by solving offline optimization problems using actual wind energy data spanning from 2008 to 2022 and real load data.} These datasets train four SML models: DT, KNN, SVM, and NB. The accuracy of each method in filtering non-active constraints and predicting the optimal binary variable values is summarized in Table \ref{tab:SML_accuracy}. The DT model demonstrates the highest accuracy at 86.8\%, making it the preferred choice for integration into the ACI-IVP-based surrogate transformation method. 
All simulations are performed in the MATLAB R2019b environment on a system with an Intel Core i7-9700 CPU running at 3 GHz and 16 GB of RAM.
\begin{table}[!t]
      % \vspace{-.6cm}
\centering
\caption{Accuracy of SML Methods for integration into ACI-IVP}
\vspace{-.2cm}
\label{tab:SML_accuracy}
\footnotesize
\begin{tabular}{ccccc}
\hline\hline
Supervised Learning Method & DT & KNN & SVM & NB \\ \hline
Accuracy (\%) & \textbf{86.8} & 63.2 & 68.4 & 35.4 \\ 
\hline\hline
\end{tabular}
\vspace{-.5cm}
\end{table}

\subsection{Evaluating the Solution Time}

The ACI-IVP-based method significantly improves the efficiency of solving W2H-CIs optimization problems compared to conventional methods. Conventional approaches, which involve directly solving the MICP using standard optimization solvers, can take over 15 minutes to find a solution. In contrast, the ACI-IVP-based method simplifies the original problem into a surrogate model, allowing for real-time solutions in just a few seconds. Table \ref{tab:soltime} highlights this difference, showing that the ACI-IVP-based method completes the process in seconds.

\begin{table}[b!]
      \vspace{-.5cm}
\centering
\caption{Solution Time of Optimization Methods}
\vspace{-.2cm}
\label{tab:soltime}
\footnotesize
\begin{tabular}{cccc}
\hline\hline
\multicolumn{1}{c}{\multirow{2}{*}{Case Study}}  & \multirow{2}{*}{Time} & \multicolumn{2}{c}{Solution Time (s)}                 \\ \cline{3-4} 
       &      &  \multicolumn{1}{c}{Conventional Method} & ACI-IVP Method \\ \hline \hline
\multirow{3}{*}{First Case}  &00:05 & \multicolumn{1}{c}{592.70}            & 1.07   \\ 
&00:10  & \multicolumn{1}{c}{562.99}      &1.19       \\ 
&00:15  & \multicolumn{1}{c}{984.50}      &1.38      \\ 
\hline 
\multirow{3}{*}{Second Case}  &00:05 & \multicolumn{1}{c}{843.28}            & 2.68   \\ 
&00:10  & \multicolumn{1}{c}{903.63}      &2.52       \\ 
&00:15  & \multicolumn{1}{c}{1017.71}      &3.17      \\ 
\hline \hline
\end{tabular}
% \vspace{-.5cm}
\end{table}

\subsection{Carbon Emission and Cost Analysis of the W2H-CIs}
\begin{figure}[!t]  
    % \vspace{-.7cm}
  \centering
\includegraphics[width=0.375\textwidth]{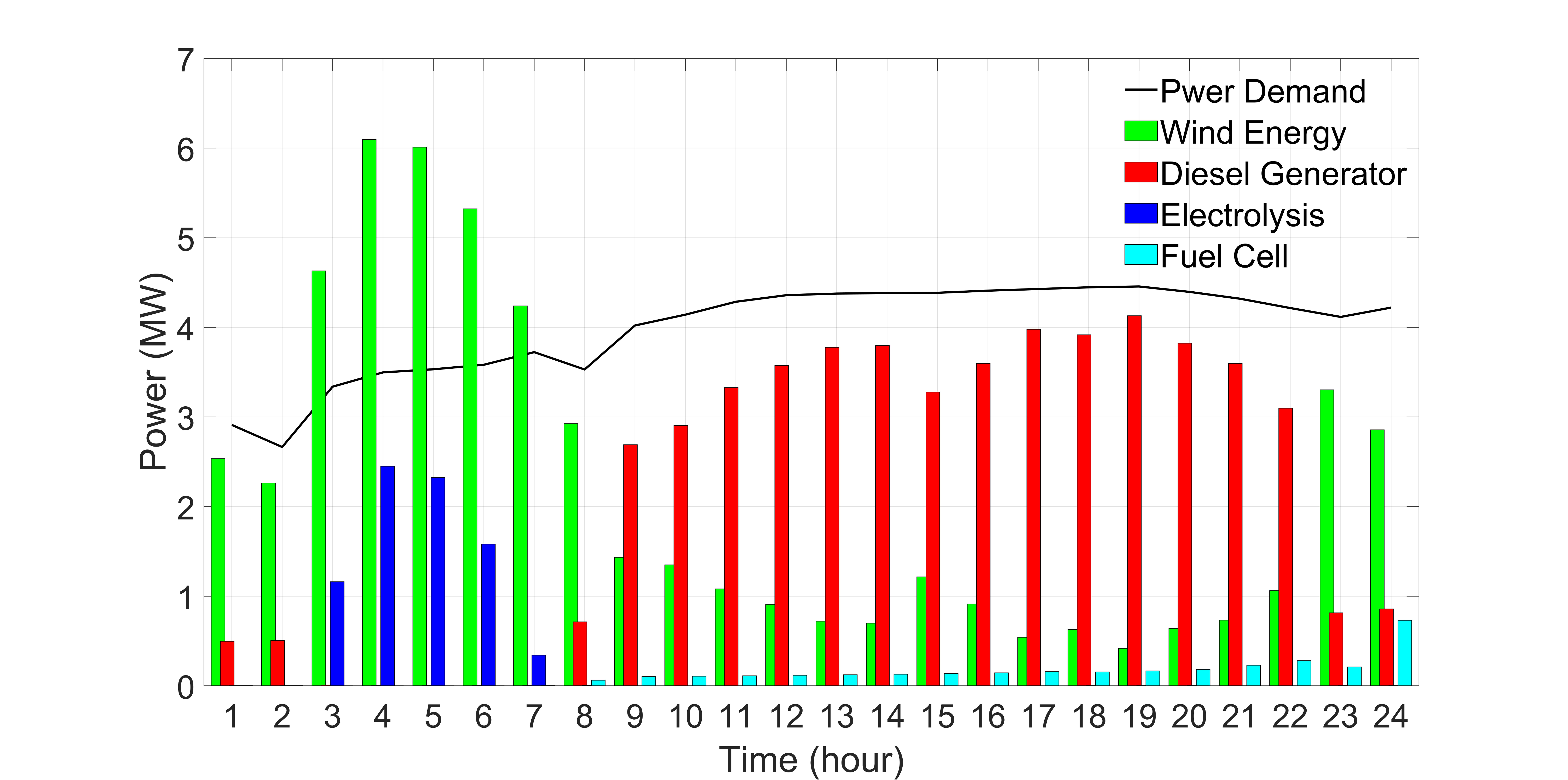}
 \centering
 \vspace{-.3cm}
    \caption{Optimal operation of W2H-CIs for the second case study}
  \label{pic: operation}   
      \vspace{-.3cm}
\end{figure}

\begin{figure}[!t]  
    % \vspace{-.5cm}
  \centering
\includegraphics[width=0.375\textwidth]{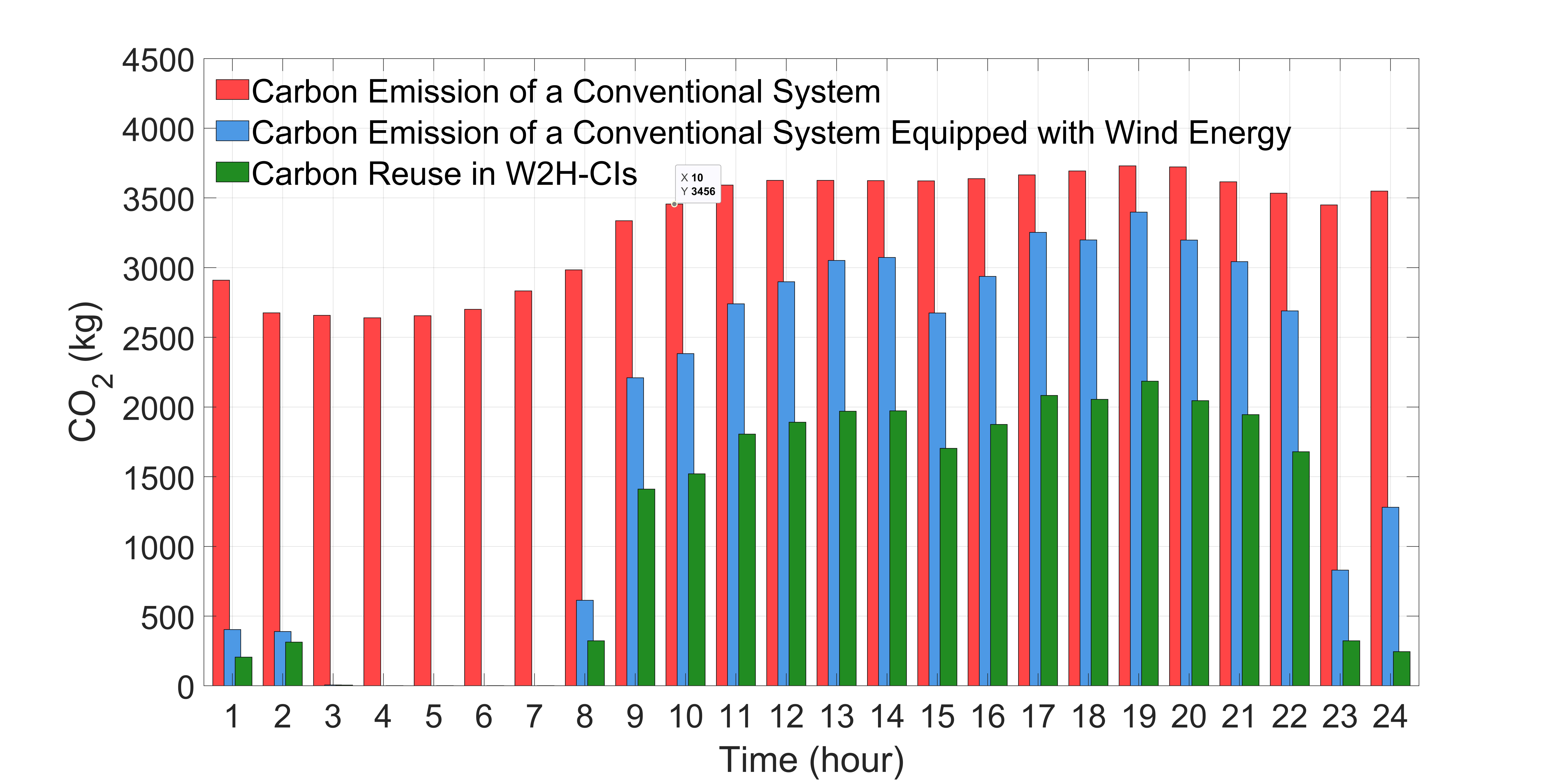}
 \centering
 \vspace{-.3cm}
    \caption{Carbon Emissions Comparison for Three Systems (Case Study 2).}
  \label{pic: Emission}   
      \vspace{-.5cm}
\end{figure}

The evaluation of the proposed W2H-CIs model demonstrates significant reductions in carbon emissions, positioning the W2H-CIs as a viable solution for achieving net-zero energy. For example, our analysis shows that the emissions from the existing system, which relies solely on diesel generators, amount to approximately 79.53 metric tons of carbon emissions daily for the second case study. When these systems are equipped with wind energy, the emissions are reduced to 44.27 metric tons. Remarkably, the W2H-CIs eliminate all carbon emissions, despite using a diesel generator as a backup source (as shown in in Fig. \ref{pic: operation}) that emits 27.55 metric tons of carbon. 
This emission is effectively captured and reused, enabling full carbon reuse. Fig. \ref{pic: Emission} illustrates the comparative analysis of carbon emissions between the conventional system, the wind-equipped system, and the W2H-CIs. 
Additionally, while the daily operational cost of the conventional system equipped with wind energy is \$56,018, the W2H-CIs reduce this cost to \$52,873. Consequently, the W2H-CIs not only achieve net-zero energy but also prove to be cost-effective, underscoring their sustainability and economic viability.  

\section{Conclusions and Future Work}\label{sec: Conclusions}
This paper introduces wind-to-hydrogen-driven critical infrastructure (W2H-CIs) as a comprehensive approach to achieving net-zero energy. The proposed W2H-CIs framework decarbonizes and enhances the resilience of critical infrastructures, including power, water, hydrogen, and transportation. To address the intermittency of wind energy, the active constraints identification and integer variable prediction (ACI-IVP) method is presented as an efficient approach to solving complex mixed-integer optimization problems in real-time. This ML-based method reduces computational complexity by utilizing a decision tree to filter out non-active constraints, replace binary variables with their optimal values, and transform the original convex mixed-integer problem into a continuous one that can be solved rapidly. The effectiveness of this method is validated through two case studies, where the ACI-IVP-based method reduces solution times from 10–15 minutes (with conventional methods) to just a few seconds, enabling real-time operation. The W2H-CIs framework offers a sustainable, adaptable, and cost-effective pathway to decarbonization, supporting global efforts toward net-zero energy.

In future work, we plan to integrate traffic models of the transportation network to enhance the accuracy of the results. Additionally, we will investigate contingencies across power, water, transportation, and hydrogen systems to ensure secure and resilient operations under various conditions. Further research will also focus on improving the data-driven approach to increase the accuracy of the ACI-IVP method during the filtering phase, thereby enhancing overall computational performance and reliability.

\bibliographystyle{IEEEtran}	
\bibliography{Main}

\end{document}